\input psfig.tex
\documentstyle[prd,aps,preprint]{revtex}
\newcommand{\be}{\begin{equation}}
\newcommand{\ee}{\end{equation}}
\newcommand{\ba}{\begin{eqnarray}}
\newcommand{\ea}{\end{eqnarray}}
\bibstyle{unsrt}
\tighten
\begin{document}
\draft
\preprint{Imperial/TP/95-96/27}
\title{Cosmic string formation and
the power spectrum of field configurations}
\author{James Robinson
\footnote{email: \tt{j.h.robinson@ic.ac.uk}}
and Andrew Yates
\footnote{email: \tt{a.yates@ic.ac.uk}}
}
\address{Theoretical Physics Group,
Blackett Laboratory, Imperial College, South Kensington,
London SW7 2BZ, U.K.}
\maketitle
\begin{abstract}
We examine the statistical properties of defects formed by the
breaking of a U(1) symmetry when the Higgs field
has a power spectrum $P(k) \propto k^n$.
We find a marked dependence of the amount of infinite string on the
spectral index $n$ and empirically identify an analytic form
for this quantity. We also confirm that this result is robust to
changes in the definition of infinite string. It is possible that this
result could account for the apparent absence of
infinite string in recent lattice-free simulations.
\end{abstract}

\section{Introduction}
Cosmic strings formed at a primordial phase transition are
strong candidates as seeds for large-scale structure in the universe
and anisotropies in the microwave background \cite{Rev,Vil}. Firm
predictions are difficult to extract from string models,
due to both the uncertainty in the statistical properties of
the string network at formation and the non-linearity of network
evolution. Information from numerical simulations has been the main
tool for calculations to date.
The observational consequences of the string scenario may be strongly
dependent on whether,
in addition to a scale-invariant distribution of loops,
any infinite string exists.
The presence or absence of infinite string may lead the string network
into significantly different scaling solutions \cite{Ped},
and in particular to solutions that
studies of large-scale structure will be able to distinguish between.

Lattice-based simulations have traditionally been employed to predict
the initial fraction as infinite string, although recent work has used a
lattice-free method to simulate string formation in a first order
phase transition \cite{Julian}.
The simplest approach, however, is to assume that
the string-forming field takes up random values at each site on a
regular lattice, each representing a causally disconnected volume, and
look for strings at each face of the lattice. Vachaspati and Vilenkin \cite{VV}
first performed this on a cubic lattice and found that approximately
80\% of string was in `infinite' form. Further work on a tetrahedral
lattice by Hindmarsh and Strobl 
demonstrated a slightly lower figure \cite{MarkKarl}.
Furthermore, Vachaspati \cite{V} has shown that for theories without
an exact U(1) symmetry (i.e. one field orientation is statistically
favoured) it is possible to reduce the fraction as infinite string to
zero. Another approach \cite{KibbleYates}, retaining the symmetry of
the theory but introducing a variation in
the ranges over which the field is correlated by means of a simple
domain-laying algorithm, showed a dependence of the fraction as
infinite string on the variance of domain volume.
Here we present a new single-parameter method for varying the scale
over which the string
field is correlated to investigate the behaviour of the amount of
infinite string present. It is a controlled means of varying the
string-forming field configuration and allows us to employ analytic
results concerning defect densities. In addition, the method makes it
possible to explore a range of physically-motivated theories in which
string-like defects arise \cite{RayGlyk}, and in
particular the role of causality in determining the properties of a cosmic
string network on very large (super-horizon) scales.

In section 2 we discuss our method for generating string
configurations and present our basic results regarding the fraction as
infinite string. In section 3 we compare the total density of string to
a theoretical prediction of the density of zeroes of the
string-forming field. In section 4, we suggest a functional form for
the relation between the  power spectrum of the string field and the
density of infinite string. In section 5 we confirm that our results
are insensitive to changes in the definition of infinite string, and
section 6 presents our conclusions.

\section{The method}
We construct our Higgs field $\phi({\bf x})$,
on a lattice with periodic boundary conditions, as a realisation of
gaussian random field with a simple power spectrum $P(k) \propto k^n$, such
that
\[
<\tilde{\phi}({\bf k})\tilde{\phi}^*({\bf k'})> = (2\pi)^3 P(k)
\delta^{(3)}({\bf k} - {\bf k'}),
\]
where
\[
\tilde{\phi}({\bf k}) = \frac{1}{(2\pi)^{3/2}}\int {\rm d}^3 {\bf x}\,
e^{-i{\bf k \cdot x}} \phi({\bf x}).
\]
The real and imaginary parts of each
Fourier component $\tilde{\phi}({\bf k})$ are assigned values
chosen from a gaussian distribution with mean zero and variance
$P(k)/2$. The phase of the field
$\phi({\bf x})$ is extracted at each lattice site and windings are
located. Throwing away the modulus of the field only affects the power
spectrum on scales larger than the correlation
length by a multiplicative factor.
\footnote{This can be understood
qualitatively as follows. The moduli of the $\phi({\bf x})$ will
be Rayleigh distributed with a finite mean.
If we rebuild $\tilde\phi({\bf k})$
from just the phase of $\phi({\bf x})$, it is analogous to performing
a walk on the complex plane with a fixed step length. This
will give a result for  $|\tilde\phi({\bf k})|$ that differs from that
obtained by a walk with a variance in the step length by a
multiplicative factor, independent of ${\bf k}$.}
Strings are paired up at random within each lattice site and
the lengths of the string loops determined. The boundary
conditions ensure that all string is closed.

We find that, as
expected, there is a distribution of loops with a characteristic
power-law density,
\[
\rho_{\rm loop}(l) \propto l^{-5/2},
\]
where $l$ is the length of a loop,
which can be derived on dimensional grounds, assuming that the
statistical properties of the loop distribution are independent of
scale \cite{VV}. 

In addition we find a number of much longer loops, winding around
the lattice several times. On a periodic lattice, it can be shown
\cite{RayEd} that these strings belong to a distribution whose
density falls off as $l^{-1}$ and are an artifact of the boundary
conditions. It is commonly believed that these represent a separate
population of `infinite' strings --- henceforth we use the word
`infinite' to refer to strings  which would never self-intersect in the
infinite-volume limit. 

The standard method for distinguishing
between the loop distribution and the infinite string has been to
introduce a cut-off in length that is of order $N^2$, where $N$ is the
size of the box in lattice units. The reason for choosing such a
cut-off is that string greater than this size is likely to
be ``topological'', that is, to traverse opposite boundaries of the simulation.
All loops longer than the cut-off
are then classified as infinite. Given this standard definition of infinite string,
figure 1 shows the variation of the infinite string fraction $f_{\infty}$ with
spectral index n.
\begin{figure}
\centerline{\psfig{file=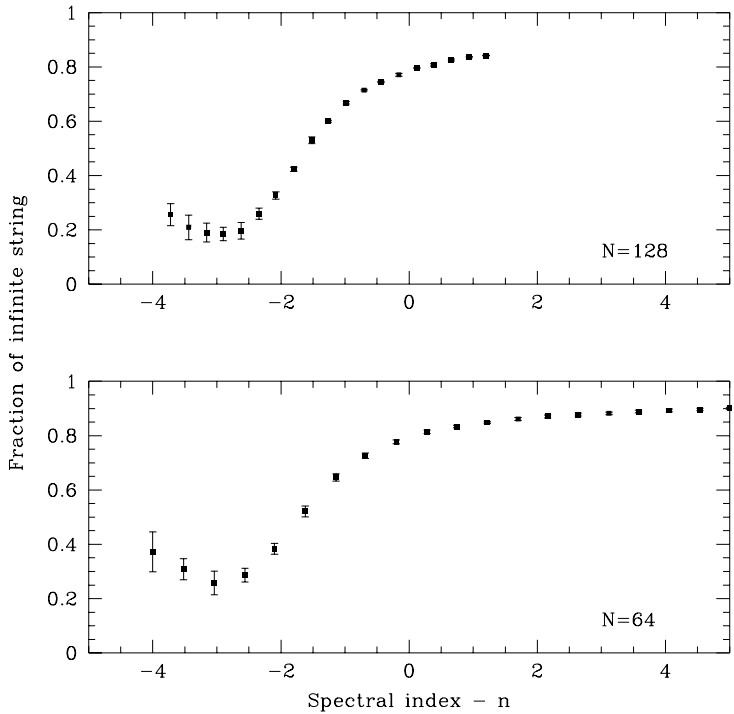,width=4in}}
\label{fig:sax}
\caption{Plot of the fraction as infinite string against spectral
index, averaged over 10 realisations for the lattice of size $N=64$
and 4 for $N=128$, employing the standard
definition of infinite string as loops longer than ${\cal O} (N^2)$.
Errorbars are shown at the $3\sigma$ level.}
\end{figure}

\section{Dependence of total string density on spectral index}
We have presented the basic findings regarding the variation of the
amount of infinite string with spectral index in figure 1. Without any
analytical explanation of the form of the curves, we can
however gain insight into some of their features by considering an
analytical result for the density of zeroes $\rho_{\rm t}$ on a two dimensional slice
through our field.
Halperin \cite{Halp} shows that
\be
\rho_{\rm t} = \left| \frac{W(0)''}{W(0)}\right|,
\ee
where  $W({\bf x})$ is the correlation function, defined as
\[
W({\bf x}) = \,< \phi({\bf x}) \phi^*(0)>.
\]
This result gives us the number density of string crossings in a
plane. Naively, by extension to three dimensions and knowing that our
string segments lie only in the three principal directions, we can
infer that the total length of string in the box is proportional to $\rho_{\rm t}$.

The relationship between the correlation function and the power
spectrum is easily derived:
\begin{eqnarray*}
W({\bf x})  &  =  &  < \frac{1}{(2\pi)^3} \int {\rm d}^3 {\bf k}\,{\rm d}^3 {\bf k'}\,
\tilde{\phi}({\bf k}) \tilde{\phi}^*({\bf k'}) e^{i{\bf k \cdot x}} > \\
& = & \int {\rm d}^3 {\bf k}\, e^{i{\bf k \cdot x}} P(k) \\
& = & 4\pi \int  {\rm d} k \, k^2 \, \frac{\sin{kx}}{kx}\, P(k).
\end{eqnarray*}
For power law spectra $P(k) = k^n$, we have
\[
W(x) = 4\pi \int  {\rm d} k \,\frac{\sin{kx}}{kx}\,k^{n+2}.
\]
We can extract the spatial dependence with the substitution $\kappa =
kx$. Then
\[
W(x) = 4\pi \int  {\rm d} \kappa \, \frac{\kappa^{n+2}}{x^{n+3}}
 \frac{\sin{\kappa}}{\kappa} = f(n) x^{-(n+3)}.
\]
We can see that a spectral index of $-3$ implies a spatially homogeneous
correlation function and thus
a totally uniform field. Any value less than $-3$  gives rise to a
correlation function divergent at large $x$, which is unphysical. 

Increasing $n$ decreases the long-range correlation with respect to
short-range. The $n=0$ case corresponds to white noise --- $\phi({\bf
x})$ is totally uncorrelated and is random at each lattice point.
This corresponds to the original scenario considered by
Vachaspati and Vilenkin, and leads to approximately $80\%$ `infinite'
string, which is confirmed in figure 1.

In practice, we
construct our field with a number of momentum modes ranging from
$k_{min} = 2 \pi/N$ to $k_{max} = 2 \pi$. From our form for the
correlation function $W$, we obtain an expression for the defect
density
\begin{eqnarray}
\rho_{\rm t} & \simeq & \frac{\int_{k_{\rm min}}^{k_{\rm max}}  {\rm d} k
\,  k^{n+4}}{\int_{k_{\rm min}}^{k_{\rm max}}  {\rm d} k \, k^{n+2}}
\nonumber \\
 & = & \left( \frac{k_{\rm max}^{n+5} - k_{\rm min}^{n+5}}{k_{\rm
max}^{n+3} - k_{\rm min}^{n+3}} \right) \left( \frac{n+3}{n+5} \right).
\end{eqnarray}
This function is plotted against $n$ for various values of $k_{\rm max}/
k_{\rm min}$ in figure 2. It can be seen that in the
infinite-volume limit, the total defect density drops to zero for values of
$n$ lower than $-3$, as expected. The presence of string at this value
of n and below is a finite size effect, and consequently the form of every plot
presented in this paper for $n<-3$ is not significant.
\begin{figure}
\centerline{\psfig{file=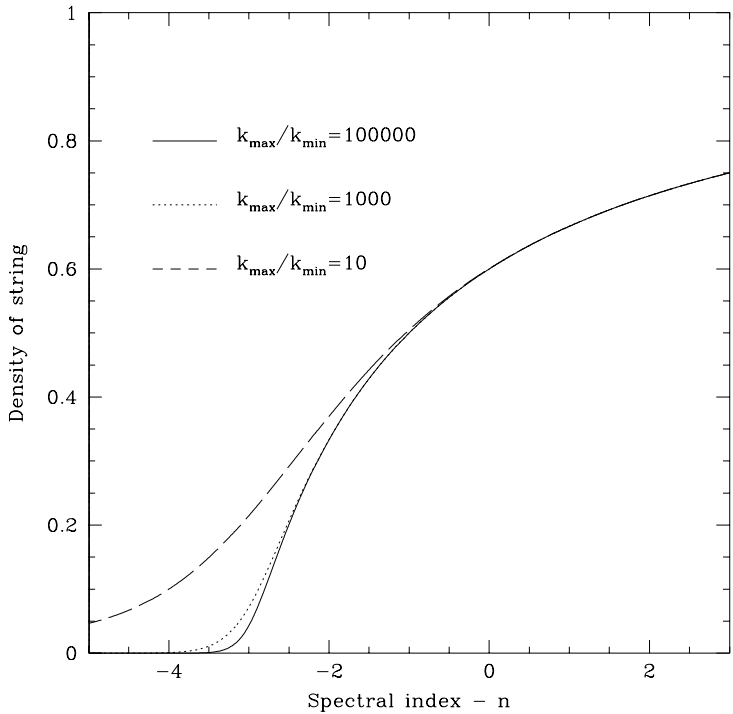,width=4in}}
\label{fig:cutoff}
\caption{The dependence of $\rho_{\rm t}$ on $n$, where $\rho_{\rm t}$ is plotted in
units of $(2\pi \times \mbox{lattice spacing})^{-2}$.}
\end{figure}

When we normalise the expression above to the data from the simulations, we
obtain the forms shown in figure 3, which are in reasonable
agreement. We note, however, that there is no simple relationship
between Halperin's result for the density of zeroes in the continuum
limit and our estimation of the total length of string, assumed to lie on the
dual lattice. It is not surprising that there are small discrepancies
between the prediction and the data. 
\begin{figure}
\centerline{\psfig{file=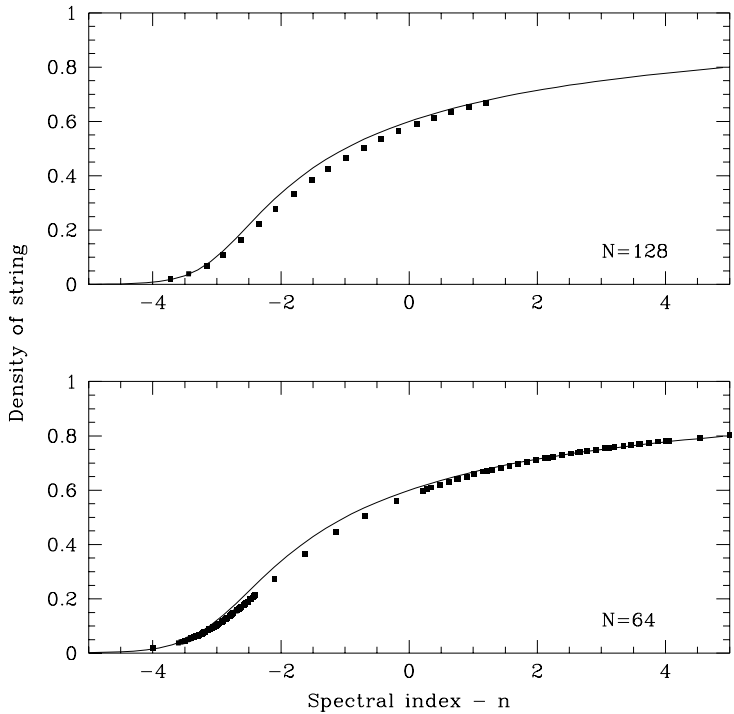,width=4in}}
\label{fig:hal}
\caption{The points show the total density of string,
again averaged over 10 realisations in the $N=64$ case and 4
in the $N=128$. The curves show the analytic prediction, normalised by
hand. Density is in arbitrary units. Errorbars are too small to appear
in the plot.}
\end{figure}
\section{Dependence of infinite string density on Spectral Index}
In figure 4 we plot the amount of string in infinite form and in
loops. We then make the observation that the former curve
can be fitted extremely well with a modified version of the Halperin
result for total defect density, plotted in figure 2,
by replacing $n$ with $n-1$ in equation 2. That is,
\be
\rho_\infty = \left( \frac{k_{\rm max}^{n+4} - k_{\rm min}^{n+4}}{k_{\rm
max}^{n+2} - k_{\rm min}^{n+2}} \right) \left( \frac{n+2}{n+4} \right).
\ee
Figure 4 shows that the expression gives the correct dependence of
$\rho_\infty$ on $n$.
The volume-dependence of $\rho_\infty$ at a given value of $n$
is confirmed for the cases $n=-3$ and $n=-1.5$ in figure 5. In the
$n=-3$ case, we can clearly see the density of infinite string vanishing
in the infinite volume limit.
Note that we introduce the volume of the
lattice through the relation 
\[
\frac{k_{\rm max}}{k_{\rm min}} = N.
\]
In the context of our definition of the correlation function, equation 2
may be rewritten as
\ba
 \rho_\infty & = & \left|\frac{\int {\rm d}k\, k^3 P(k)}{\int {\rm d}k \,k
P(k)} \right| \nonumber \\
& = & \left|\frac{C''(0)}{C(0)} \right|,
\ea
here cast in the same form as the Halperin result (equation 1), with
 $C(x)$ satisfying
\[
\frac{{\rm d}C}{{\rm d}x} = W(x).
\]
There is  no direct evidence that equation 4 holds for a general
$P(k)$ --- this is the subject of work in progress \cite{Jamesetc}
It has two interesting properties, however, that support
this suggestion. We would expect that any
expression relating to infinite string would be non-local, in that it
would have to take account of global properties of the field
correlation function. This indeed is the case --- the denominator
involves the integral of the correlation function $W(x)$. Secondly, it
is straightforward to confirm that for any power spectrum $P(k) \ge 0$
for all $k$,  and using the expression above for $\rho_\infty$,
\[
0 \le \frac{\rho_\infty}{\rho_{\rm t}} \le 1.
\]

We can then calculate the amount of
string that we expect to find in loops by subtracting this modified
Halperin result from the original expression for the total string density;
\[
\rho_{\rm loops} = \rho_{\rm t} - \rho_\infty.
\]
We see that the basic features of the loop distribution are
reproduced by this expression.  The major discrepancy exists for
positive values of $n$, where the observed loop density tends to a
finite value whilst the predicted density falls to zero. This is not
entirely surprising: one artifact of working on a cubic
lattice is that a given distribution of phases does not uniquely
determine the length distribution of strings. In cells where
four or six ends of string meet, we must make some choice about how to
pair these ends up, and different choices will give rise to different
amounts of infinite string and loops. This non-uniqueness, which we
reiterate is purely a consequence of the cubic lattice, will become
more important at high values of $n$.  There the
lattice becomes virtually filled with string, and cells within which
the pairings are undetermined are more frequent. There will always be
a non-zero probability of forming a loop of string in a full lattice
when a random pairing scheme is implemented.

\begin{figure}
\centerline{\psfig{file=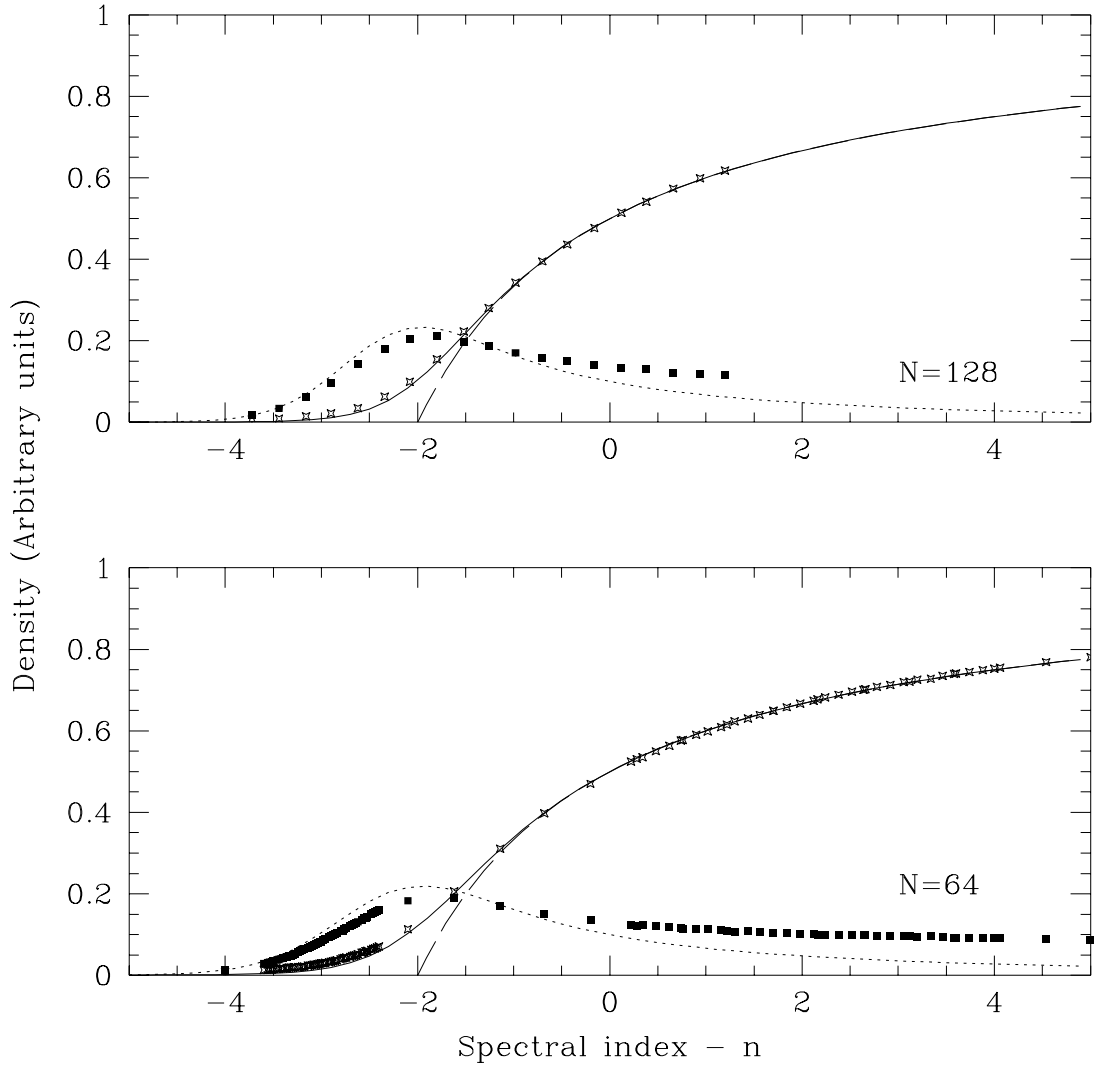,width=6in}}
\label{fig:log2}
\caption{The data points show numerical results for the density of
infinite string $\rho_\infty$ (crosses) and loops $\rho_{\rm loop}$
(solid squares). The solid and short-dashed
curves show empirical fits to this data. The long-dashed line shows
our analytic prediction for $\rho_\infty$ in the infinite-volume limit.}
\end{figure}
\begin{figure}
\centerline{\psfig{file=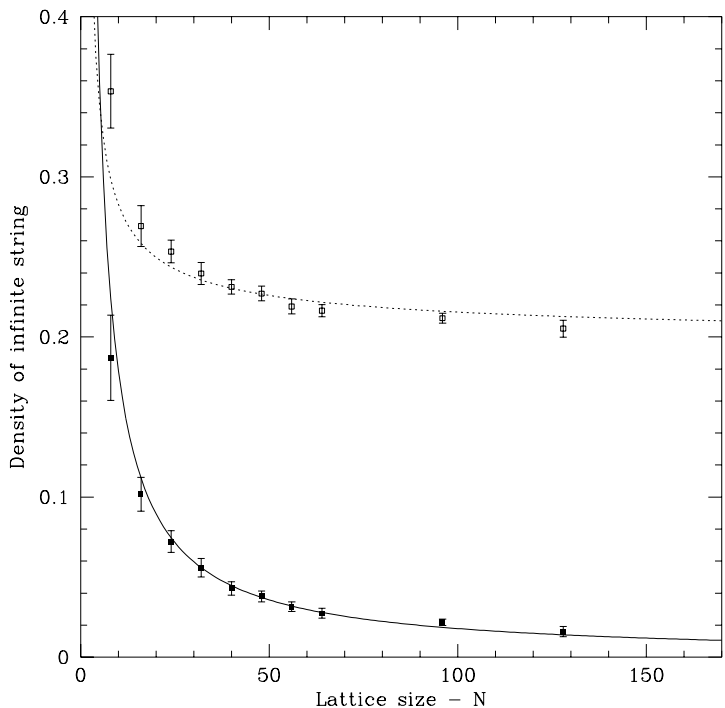,width=4in}}
\label{fig:scaling}
\caption{The lower set of points show the density of infinite string
at $n=-3$ for varying lattice sizes $N$. The upper set correspond to
$n=-1.5$. The fitted curves, which are normalised individually by
hand, show the prediction from equation 3.}
\end{figure}

If we are to believe the empirically-identified
analytic form for $\rho_\infty$,
we now see that the existence of {\it infinite} string for values
of $n$ below $-2$ is purely a finite-volume effect. In the
infinite-volume limit, at $n=-2$ and below there will be no
infinite string, as shown clearly by the long-dashed curve
in figure 4. The loop density will develop a discontinuity
in its derivative with respect to $n$ at $n=-2$.

As we reduce the string density we find that the infinite string
density $\rho_\infty$ also decreases, and will go to zero at some
non-zero value of the total string density $\rho_{\rm t}$.
This may well be analogous to the findings of Vachaspati \cite{V}, in
which he observes a similar decrease in $\rho_\infty$
with $\rho_{\rm t}$, and a phase transition in which the infinite string vanishes 
at a critical and non-zero value of the string-forming probability. In
addition he finds that the loop density also peaks sharply around the
critical density. This similarity is particularly interesting as the
means of varying the string density is fundamentally different to the
present case --- we preserve the underlying symmetry of the
string-forming field, whereas Vachaspati introduces an explicit
breaking of the U(1) symmetry.

Further, we can now explain the upturn in the plots 
of the variation
of $f_\infty$ with $n$  that occurs below $n=-3$, as shown in figure
1. Using 
\[
f_\infty  = \frac{\rho_\infty}{\rho_{\rm t}},
\]
the result is shown in figure 6.
\begin{figure}
\centerline{\psfig{file=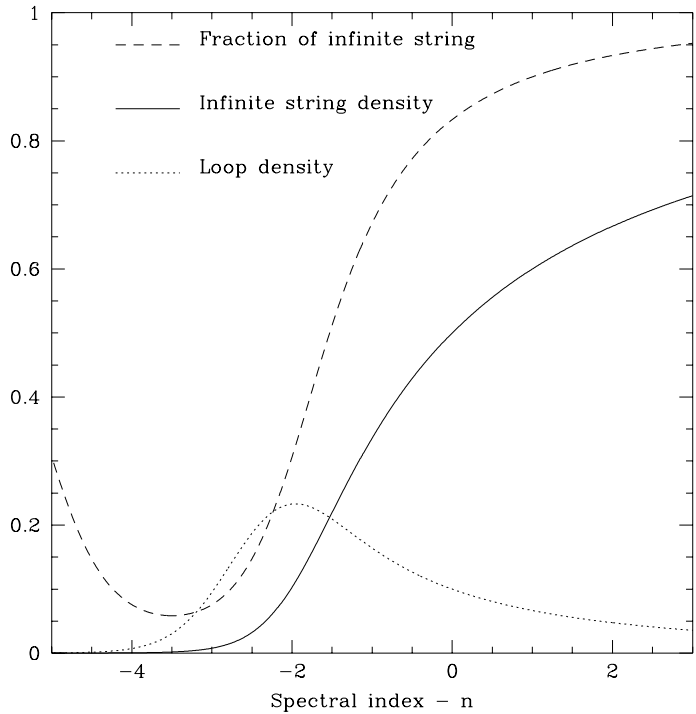,width=4in}}
\label{fig:predict_sax}
\caption{Prediction for the infinite string fraction using the
empirical forms for the infinite string and loop densities, assuming a
box size of $N=128$.}
\end{figure}
\section{The definition of infinite string}
We have found that the fraction $f_{\infty}$ of infinite string,
defined as loops of
length $>{\cal O}(N^2)$, is strongly dependent on the spectral index
$n$. This relationship could be a genuine indication of variation of
$f_{\infty}$ with $n$.  However,  
we have also shown that $f_{\infty}$ is a monotonically increasing function of
string density for $n$ greater than $-3$. We must therefore
investigate the robustness of our definition of infinite string to
variations in string density before concluding that there is a
fundamental dependence of $f_{\infty}$ on $n$.

In the Vachaspati-Vilenkin ($n=0$) case,
the strings perform approximately Brownian
random walks with a step length $\xi$ of one lattice unit. A
topological string, crossing
a lattice of side $N$, will thus have a typical minimum length $\sim N^2$.
As we decrease the spectral index,
introducing more correlations on larger scales, we smooth the field
and thus the string network over more lattice spacings. The step
length $\xi$ is effectively increased and
length of a string which crosses the box is reduced. 
If we define infinite string by means of a  cut-off $L$,
 where $L$ is the length of a string that is likely
to traverse the box, then we have
\[
 L = \xi \left( \frac{N^2}{\xi^2} \right) \propto \frac{1}{\xi}.
\] 
This suggests that the cut-off should be varied to take account of the
defect density and thus $\xi$.
In figure 7 we plot the fraction as string present in loops less than
a certain size $L$ for various values of the spectral index. Each of
these curves exhibits a distinct turnover, which corresponds to the
onset of topological string.  We see
that the value of $L$ at this point does indeed decrease with $n$, and
so with total string density.
Consequently, by choosing our cut-off in the naive density-independent
way we are introducing a systematic error which will undercount long
string for lower values of $n$.
\begin{figure}
\label{fig:fits}
\centerline{\psfig{file=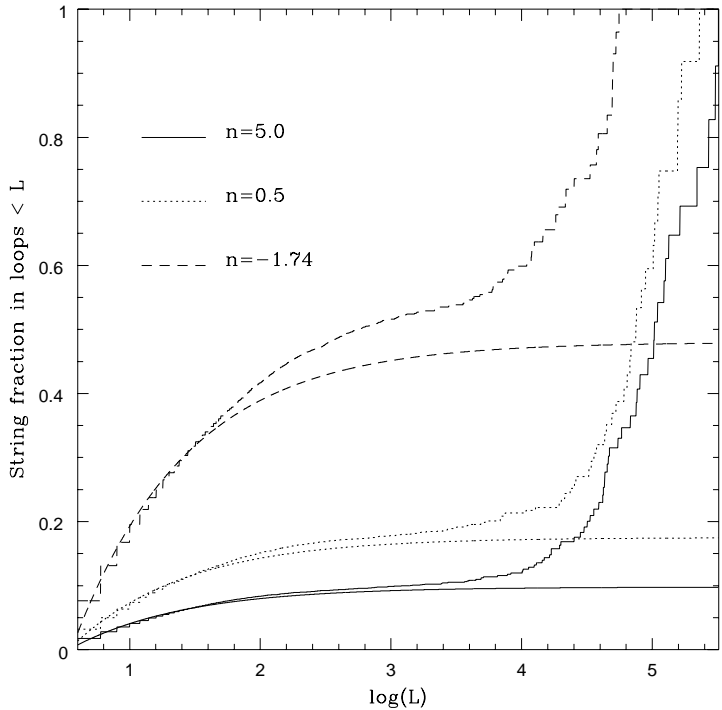,width=4in}}
\caption{The cumulative loop
distribution (jagged lines) for three values of the spectral index, for an $N=64$ lattice.
The curves show the contribution to the total amount of string we
would expect from a scale-invariant distribution of loops. They were
fitted to the data for loops
of length less than a limit $N$, below which we expected the $l^{-5/2}$
behaviour to hold true. The final results were insensitive to this
limit --- a value of $N^2$ resulted in an almost identical plot.}
\end{figure}

An alternative scheme \cite{Julian} for estimating the fraction as
infinite string is as follows. Let the total length of string in the box
be $T$.  We know that a scale-invariant distribution of loops exists
at small loop length $l$.
We can then calculate the total length $T_{\rm loop}(L)$ we
would expect purely from this distribution extrapolated to an
arbitrary length $L$. We have 
\begin{eqnarray*}
T_{\rm loop}(L) & = & \int_{l_{\rm min}}^L l \rho_{\rm loop} \, {\rm d}l \\
& = & \int_{l_{\rm min}}^L Al^{-3/2} \, {\rm d}l \\
  & = & 2A (l_{\rm min}^{-1/2} - L^{-1/2}) .
\end{eqnarray*}
$A$ and $l_{\rm min}$ are obtained by a least-squares fit to data at
small $L$. The fitted curves in figure 7 show the contribution of
$T_{\rm loop}(L)$ to the total amount of string.
The fraction as infinite string is then defined as
\[
f_{\infty} = 1 -\frac{T_{\rm loop}(\infty)}{T}.
\]

The revised form for the plot of the fraction as infinite string against
spectral index is shown in figure 8. This is now free of any
systematic effects due to variation in overall string density. We see
that the correction to the fraction as infinite string is small, and
conclude that there is a genuine dependence of $f_{\infty}$ on $n$.
\begin{figure}
\label{fig:frac}
\centerline{\psfig{file=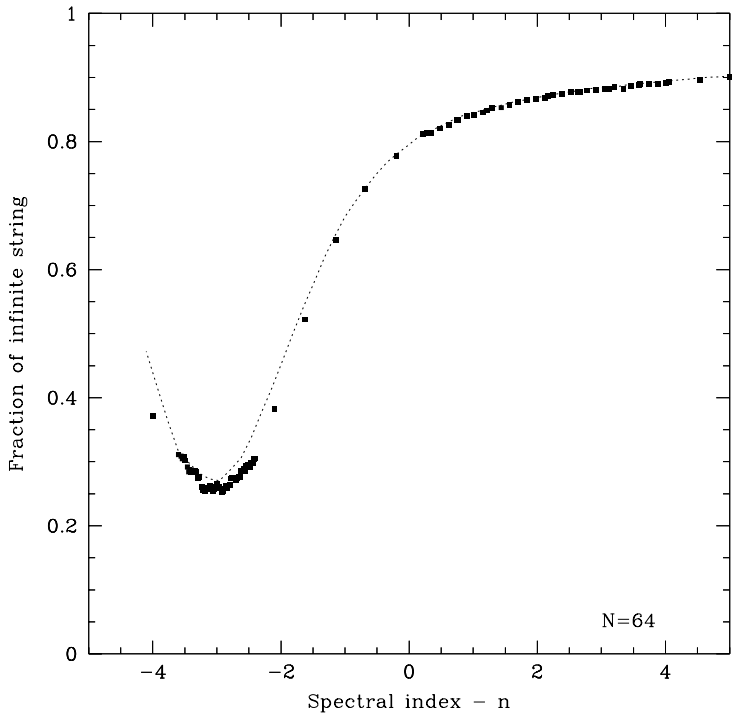,width=4in}}
\caption{The dotted line shows the result of using the alternative
definition of infinite string. The points are repeated from figure 1.}
\end{figure}
\section{Conclusion}
We have demonstrated a strong dependence of the fraction as infinite
string formed during a $U(1)$ symmetry breaking phase transition on
the power spectrum of the Higgs-field configuration. For a power
spectrum $P(k)\propto k^n$, we find agreement with an analytic
formula for variation of the total density of string with $n$. We find
a strikingly simple closed-form expression which fits the density of
infinite string and is closely related to the analytic expression for
the total string density.
We find that the fraction as infinite string falls with
$n$, and vanishes at $n=-2$ if the
formula for the density of infinite string is to be believed.
Between $n=-2$ and $n=-3$, string only exists in loop form, and below
$n=-3$ the total string density is zero.
We have also shown that the
dependence of the fraction as infinite string with $n$ is robust to
reasonable changes in our definition of infinite string.

For $n=0$, the value of the Higgs field at each lattice site is
independent, and as expected we reproduce the Vachaspati-Vilenkin
result of approximately 80\% infinite string \cite{VV}. In a cosmological
phase transition,
we expect values of the Higgs field to be completely independent on scales
larger than the causal horizon
\cite{kibble}. Such a field configuration has a spectral index
$n=0$ on the largest scales.
Provided string we classify as infinite in our lattice simulation
does indeed correspond to infinite string in the continuum
limit, we can confirm that a non-zero fraction as infinite
string will be formed in a cosmological phase transition. The
existence of this population of infinite string may have important
consequences for the
scaling solution of the evolving network, and hence for the
observational consequences of cosmic string.

Our findings show that if we want to compute the fraction as infinite
string in the universe correctly using
numerical simulations, the field configuration must have the
correct $n=0$ form on the largest scales. Failure to fulfil this
condition could severely under- or over-estimate the fraction as
infinite string.
In particular, the results of 
recent lattice-free simulations of a first-order phase
transition \cite{Julian} appear to be consistent with an
infinite string fraction equal to zero. It is important to ensure that
on large scales the power spectrum of the field configuration does have a
spectral index $n=0$, before applying this result in a cosmological context.

\section*{Acknowledgements}
We would like to thank A. Albrecht, N. Antunes, L. Bettencourt,
J. Borrill and T.W.B. Kibble for helpful discussions. 
J.\ R.\ is supported by PPARC Award No.\ 94313015, 
A.\ Y.\ is supported by PPARC Award No.\ 93300799, and 
this work was supported in part
by the European Commission under the Human Capital and Mobility
programme, contract no. CHRX-CT94-0423.


\begin{thebibliography}{99}
\bibitem{Rev} M.B. Hindmarsh,
T.W.B. Kibble, {\it Rep. Prog. Phys.} {\bf 58} (1995) 477
\bibitem{Vil}A. Vilenkin, P. Shellard, {\em Cosmic Strings and
other Topological Defects} (Cambridge University Press, Cambridge, 1994)
\bibitem{Ped} P. Ferreira, N. Turok (unpublished)
\bibitem{Julian} J. Borrill, hep-ph 9511295, DART-HEP-95/06 (1995)
\bibitem{VV} T. Vachaspati, A. Vilenkin, {\it Phys. Rev. D}
{\bf 30} (1984) 2036
\bibitem{MarkKarl}  M.B. Hindmarsh, K. Strobl, Nucl. Phys. {\bf B437}
(1995) 471
\bibitem{V} T. Vachaspati, {\it Phys. Rev. D} {\bf 44} (1991) 3723
\bibitem{KibbleYates} A. Yates, T.W.B. Kibble, {\it Phys. Lett B} {\bf 364},
149 (1995)
\bibitem{RayGlyk} G. Karra, R. Rivers, hep-ph/9603413,
Imperial College preprint
95-96/28. Available on the WWW from http://euclid.tp.ph.ic.ac.uk/Papers/
\bibitem{RayEd} D. Austin, E.J. Copeland, R.J. Rivers, {\it Phys. Rev. D}
{\bf 49} (1994) 4089
\bibitem{Halp} B. Halperin, in `The Statistical Mechanics of
Topological Defects', Les Houches, Session XXXV, 1980 (North Holland
Publishing Company)
\bibitem{Jamesetc} G. Karra, R. Rivers, J. Robinson (in preparation)
\bibitem{kibble} T.W.B. Kibble, {\it J. Phys. A} {\bf 9} (1976) 1387
\end{thebibliography}
\end{document}